\documentstyle[epsf,11pt]{article}
\begin{document}
\baselineskip 0.8cm

\vspace*{-2.4cm}

{\bf Classification:} Biological Sciences, Genetics\\

\vspace*{-0.5cm}

{\bf Title:} Dynamic modeling of gene expression data

{\bf Authors:}
{\sc Neal S. Holter$^1$, Amos Maritan$^2$,
Marek Cieplak$^{1,3}$,\\ Nina V. Fedoroff$^4$ \& Jayanth R.
Banavar$^1$}\\

\vspace*{0.5cm}
$^1$ Department of Physics and Center for Materials Physics,
104 Davey Laboratory, The Pennsylvania State University, University
Park, Pennsylvania 16802\\
$^2$ International School for Advanced Studies (S.I.S.S.A.),
Via Beirut 2-4, 34014 Trieste, Istituto Nazionale di
Fisica Materia \\and the Abdus Salam
International
Center for Theoretical Physics, 34014 Trieste, Italy \\
$^3$Institute of Physics, Polish Academy of Sciences, 02-668
Warsaw, Poland\\
$^4$ Department of Biology and the Life Sciences Consortium, 519
Wartik
Laboratory, The Pennsylvania
State University, University Park, Pennsylvania 16802\\

{\bf Corresponding author:} Jayanth R. Banavar, 104 Davey Laboratory, 
The Pennsylvania
State University, University Park, Pennsylvania 16802, phone: 814-863-1089,
FAX: 814-865-0978, email: jayanth@phys.psu.edu\\

 pages: 20\hskip 1cm
 figures: 3 \hskip 1cm
 tables: 5 \\
 abstract: 115 words \hskip 1cm
 paper: 40,435 characters

\baselineskip 0.8cm
\parindent 0.75cm


\newpage
ABSTRACT

\noindent
We describe the time evolution of gene expression levels by using
a time translational matrix to predict future expression levels
of genes based on their expression levels at some initial time.
We deduce the time translational matrix for previously published
DNA microarray gene expression data sets by modeling them within
a linear framework using the characteristic modes obtained by
singular value decomposition. The resulting
time translation matrix provides a measure of the 
relationships among the modes and governs their time evolution.
We show that a truncated matrix linking just a few modes
is a good approximation of the full time translation matrix.
This finding suggests
that the number of essential connections among the genes is small.

\newpage
\vspace{0.5cm}

{\it Introduction}

The development and application of DNA- and
oligonucleotide-micro-\\array techniques(1, 2) for measuring the
expression of many or all of an organism's genes has stimulated
considerable interest in using expression profiling to elucidate the
nature and connectivity of the underlying genetic regulatory
networks (3-9).  Biological systems, whether organismal or sub-organismal,
are robust, adaptable, and redundant (10).  It is
increasingly apparent that such robustness is inherent in the
evolution of networks (11).
More particularly, it is the result of the
operation of certain kinds of biochemical and genetic mechanisms
(12-18).

Analysis of global gene expression data to group genes
with similar expression patterns has already proved useful in
identifying genes that contribute to common functions and are
therefore likely to be co-regulated (19-23).  Whether information
about the underlying genetic architecture 
and regulatory interconnections can be
derived from the analysis of gene expression patterns remains to be
determined.  Both the subcellular localization and
activity of transcription factors can be influenced by
post-translational modifications and interactions with small
molecules and proteins.  These can be extremely important from
a regulatory perspective, but undetectable at the gene
expression level, complicating the identification of causal connections
among genes.   Nonetheless, a number of conceptual frameworks for
modeling genetic regulatory networks have been proposed (3--9).

Several
groups have recently applied standard matrix analysis to
large gene expression datasets, extracting dominant patterns or
``modes" of gene expression change(24-26).  It has become evident that
the complexity of gene expression patterns is low, with just a
few modes capturing many of the essential features of these patterns.
The expression pattern of any particular gene can
be represented precisely by a linear combination of the modes with
gene-specific coefficients (25). Furthermore, a good approximation of the
exact pattern can be obtained by using just a few of the modes,
underscoring the simplicity of the gene expression patterns.

In the present communication we consider a simple 
model in which the expression levels of the genes at a given time are
postulated to be linear combinations of their levels at a previous
time.  
We show that the temporal evolution of the gene expression
profiles can be described within such a linear framework by using
a ``time translation" matrix which reflects the magnitude of the
connectivities between genes and makes it possible to predict future
expression levels from initial levels.  The basic framework has been
described previously, along with initial efforts to apply the model
to actual datasets (5, 7--9). 
The number of genes, $g$, typically far exceeds the number of time points
for which data are available and this makes the problem of determining the
time translation matrix an ill-posed one.
The basic difficulty is that in order to uniquely and unambiguously
determine the $g^2$ elements of the time translation matrix, one needs a set of
$g^2$ linearly independent equations.
D'haeseleer et al. (1999) used 
a non-linear interpolation scheme
to guess the shapes of gene expression profiles
between the measured time points.  
As noted by the authors, their final results  are crucially dependent
on the precise interpolation scheme and are therefore speculative.
Van Someren et al. (9) instead chose to cluster the genes and study the
interrelationships between the clusters. 
In this situation, it is possible to determine the time translation
matrix unambiguously, provided that the clustering is meaningful.
However, most clustering algorithms are based on profile
similarity, the biological significance of which is not entirely clear.

Here we construct the time translation
matrix for the characteristic modes obtained using singular
value decomposition (SVD). The  polished expression data (22) for each 
gene may be viewed as a unit vector in a hyperspace, each of whose
axes represents the expression level at a measurement time of the
experiment. The SVD construction ensures that the modes correspond
to linearly independent basis vectors, a linear combination of which
exactly describes the expression pattern of {\em each} gene. 
Furthermore, this basis set is optimally chosen by SVD
so that the contributions of the modes progressively decrease
as one considers higher order modes (24-26).

Our results suggest
that the causal links between the modes, and thence the genes,
involve just a few essential connections. 
Any additional connections among the
genes must therefore provide redundancy in the network.
An important corrollary is that it may be impossible to determine
detailed connectivities among genes with just the microarray
data because
the number of genes greatly exceeds the number of contributing
modes.


\vspace*{0.5cm}
\newpage
{\it Methods}

It was shown recently (24-26) that the essential features of the
gene expression patterns are captured by just a few 
of the distinct characteristic modes
determined through SVD.
In the previous work (25), we treated the gene expression pattern
of all the genes as a ``static" image and derived
the underlying genome-wide characteristic modes of which it is composed. 
Here we carry out a dynamical analysis, exploring the
possible causal relationships among the genes by
deducing a time translation matrix for
the characteristic modes defined by SVD.

In order to deduce the time translation matrix, we 
consider an exact representation (25) of 
the gene expression data as a linear combination of all the $r$ 
modes obtained from SVD. Each gene is characterized by $r$ 
gene specific coefficients, where $r$ is
one less than the number of time points in the polished
data set (22).
The key goal is to attack the inverse problem and infer the
nature of the gene network connectivity. However,
the number of time points is smaller than the number of genes and thus
the problem is underdetermined. Nevertheless, the inverse problem is 
mathematically well defined and tractable if one considers the causal
relationships among the $r$ characteristic modes obtained by 
SVD. This is because, as noted earlier, the $r$ modes
form a linearly independent basis set.

Let
\begin{equation}\label{Y}
Y(t) = \left(
\begin{array}{c} X_1(t) \\ X_2(t) \\ \vdots \\ X_r(t) \end{array}
 \right)
\end{equation}
represent the expression levels of the $r$ modes at time $t$.
Then, mathematically, our linear model is expressed as
\begin{equation}\label{linear}
Y(t+\Delta t) =  M \cdot Y(t)
\end{equation}
where $M$ is a time-independent $r \times r$ time translation
matrix which provides key information on the influence of the modes
on each other.  The time step, 
$\Delta t$, is chosen to be the highest common factor among all
of the experimentally measured time intervals so that
the time of the $j^{th}$ measurement is $t_j = n_j \Delta t$,
where $n_j$ is an integer.  For equally spaced measurements, $n_j = j$.

In order to determine $M$, we define a quantity $Z(t)$ with the
initial condition
$Z(t_0) = Y(t_0)$ and, for all
subsequent times, $Z$ determined from $Z(t+\Delta t) = M \cdot Z(t)$.
For any integer $k$, we have
\begin{equation}\label{Z}
Z(t_0 + k \Delta t) = M^k \cdot Y(t_0).
\end{equation} 
The $r^2$ coefficients of $M$ are choosen to minimize the cost function
\begin{equation}\label{costfunction}
CF = \sum_j \|Y(t_j)-Z(t_j)\|^2\;/\;\sum_j\|Y(t_j)\|^2 .
\end{equation}
For equally spaced measurements, $M$ can be determined exactly using a 
linear analysis so that $CF = 0$.  For unequally spaced measurements, the
problem becomes non-linear and it is necessary to deduce $M$ using an
optimization technique such as simulated annealing (27). 
The outcome of this analysis is that
the gene expression data set can be re-expressed precisely using the
$r$ specific coefficients for each gene
(a linear combination of the $r$ modes with these coefficients gives
the gene expression profile),
the $r \times r$ time translation matrix, $M$, deduced
as described above, and the initial values of
each of the $r$ modes.

\vspace*{0.5cm}
{\it Results}

We have determined $M$, the $r\times r$ time translation matrix,
for three different data sets of gene expression
profiles: yeast cell-cycle (CDC15) (20) using the first 12 equally
spaced time points representing the first two cycles,
yeast sporulation (21) which has 7 time points and
human fibroblast (22) which has 13 time points (Table~1).
The matrix element $M_{i,j}$ describes the influence of mode $j$ on mode $i$.
Specifically, the coefficient $M_{i,j}$ multiplied by the expression level of
gene $j$ at time $t$ contributes to the expression level of gene $i$
at time $(t + \Delta t)$. A positive matrix element leads to the
$i$'th gene being positively reinforced by
the $j$'th gene expression level at a previous time.
$M$ is determined exactly and uniquely for the yeast cell-cycle data.
The unequal spacing of the time points in the two other
data sets precluded an exact solution and $M$ is an
approximation derived using simulated annealing techniques (27).
We have verified that the accuracy of $M$ is very high by showing that
the temporal evolution of the modes is reproduced well and that the 
reconstructed gene expression patterns are virtually 
indistinguishable from the experimental data.
The singular values are spread out and the amplitudes of the
modes decrease as one considers higher order modes (25).  
This fact implies that
the influence of the dominant modes on the other modes is generally small.
Interestingly, for the cdc15 and 
sporulation data sets, the converse is also true and the dominant modes are
not strongly impacted by the other modes,
especially when one takes into account the lower amplitudes of the
higher order modes.
This finding suggests that a
few-mode approximation ought to be excellent for these two cases.

Once the matrix $M$ characterizing the interrelationship between the $r$
modes is determined, it is a simple matter to deduce a
matrix that similarly describes the interactions between any other 
set of $r$ linearly independent profiles.
Specifically, one can straightforwardly determine the
interrelationships between $r$ clusters of genes.
As an example, consider the sporulation data (14) 
which is characterized by $r$=6. The problem of deriving the time translation
matrix is underdetermined if the number of clusters
exceeds six and then there is no unique solution. When the number of clusters
is less than six there is no guarantee that there exists even one solution.
We therefore consider six clusters (metabolic, early I, early II,
middle, mid-late and late), excluding the early-mid cluster
which forms the least coherent group.  
The average expression patterns of the six clusters $(c_1, \ldots ,c_6)$ 
are obtained as averages over the genes within the cluster
and can be expressed as linear
combinations of the six modes as
\begin{equation}\label{something}
  C(t) = 
  \left(
    \begin{array}{c} c_1(t) \\ c_2(t) \\ \vdots \\ c_6(t) \end{array}
  \right) = 
  S \cdot Y(t)
\end{equation}
where $S$ is a $6 \times 6$ matrix.  The rows of $S$ are the components of each of
the characteristic modes that make up the average expression pattern for the 
six clusters. 
The interrelationships between the cluster expression patterns 
is determined with a time translation matrix of the form 
\begin{equation}
  N = S \cdot M \cdot S^{-1}
\end{equation}
so that
\begin{equation}
  C(t+\Delta t) = N \cdot C(t)  \;\; .
\end{equation}
The averages of the experimental measurements (circles) and 
the predicted expression 
patterns (lines) of the six clusters are 
shown in Fig.~1 and are in excellent agreement, confirming the
accuracy of the $M$ matrix for the sporulation data in Table 1.
The matrix $N$ is
shown in Table~2. The significance of the entries in $N$ is similar to that
described earlier for $M$. That is, the matrix
element $M_{i,j}$ describes the influence of cluster $j$ on cluster $i$.
Specifically, the coefficient $M_{i,j}$ multiplied by the expression level of
cluster $j$ at time $t$ contributes to the expression level of cluster $i$
at time $(t + \Delta t)$. A positive matrix element leads to the
$i$'th cluster being positively reinforced by
the $j$'th cluster expression level at a previous time.

Does one need the full $r \times r$ time translation matrix to describe
the gene expression patterns? Or is an appropriately chosen truncated time
translation matrix adequate to reconstruct the expression
patterns with reasonable fidelity?
We now consider a linear 
interaction model (Eq. \ \ref{linear}) within which $M$ is a
$2 \times 2$ matrix and only the two most important modes are used.
The values of the four entries in the matrix $M$ are
determined using an optimization scheme that minimizes the cost
function similar to that given in Eq.~\ref{costfunction}.
The resulting $M$ matrices are shown in Table~3 and 
a comparison of the calculated modes (solid lines) with
those obtained by SVD (dashed lines)
for the three sets of gene expression profiles is shown in Fig.~2.
It is interesting to compare these $2 \times 2$ matrices
with the corresponding portion of the full matrices shown in Table~1.
The two mode approximation is excellent for the cdc15 data set (CF=0.05),
moderate for the sporulation data set (CF=0.18)
and not as good for the fibroblast data set (CF=0.31)
as for the others. 
As noted before, the use of the full $r\times r$ time translation
matrix leads to an exact reproduction of the data set. 
Not unexpectedly,  
the quality of the fit improves as the number of modes considered
is increased.
Figure 3 shows the reconstructed expression profiles starting
with the initial values and using the $2\times 2$ time
translation matrix
(denoted by a), the profiles obtained as a
linear combination of the top two modes with appropriate 
gene-specific coefficients (b) and the experimental
data (c) for the three data sets. In all three cases,
the main features of the expression patterns are reproduced
quite well by the time translation matrix with just two modes.
The 2-mode reconstruction
of the CDC15 profiles is the most accurate of the three.

It can be shown that, in general, a $2 \times 2$ time translation 
matrix produces only
two types of behavior, depending on its eigenvalues.  If the
eigenvalues are real, the generated modes will independently grow or
decay exponentially.  When the eigenvalues are complex conjugates of each
other, as they are for 
all three cases we have examined, the two generated modes
are oscillatory with growing or decaying amplitudes.  
Mathematically, the two modes are constrained to have the form:
\begin{equation}
X_1(t)\;=\;cAG^{(t/\Delta t)} 
sin(\frac{2\pi t}{\tau} + \Delta) \;\;, \\
\end{equation}
\begin{equation}
X_2(t)\;=\;cG^{(t/\Delta t)} sin(\frac{2\pi t}{\tau} + \Delta + \phi ) \;\;.
\end{equation}
Both modes are
described by a single time period, $\tau$, and a single growth or decay
factor, $G$.  Because there are four
parameters in the matrix $M$, there can only be four independent attributes
in the generated modes.  Two
other parameters, $c$ and $\Delta$, are determined from the initial conditions.
In addition to $\tau$ and $G$, we can also determine
the phase difference between the two modes, $\phi$, and the relative amplitude
of the two modes, $A$.  These attributes can be determined from
the coefficients in $M$ using the equations in Table~4.
Table~5 shows the four attributes for each of the three
data sets. The self-consistency of our analysis is underscored by the 
fact that the magnitude of the growth factor, $G$, is close 
to one for all three cases,
which is a biologically pleasing result in that the modes do not
grow explosively or decay.
For the cell cycle data, the characteristic period is about 
115 minutes. In the other two cases the data are not periodic
and hence the best-fit periods are comparable to the duration
of the measurement.
For the yeast cell cycle data, $\phi$, the phase 
difference between the top two modes is $90^\circ$, suggesting a
simple sine-cosine relationship, as noted by Alter et al. (26).
Indeed, this result is self-consistent. When $G$ is equal to 1 and an 
integer number of periods is considered, orthogonality of the top two modes
requires that the phase difference be 90$^o$.

In summary, we have shown that it is possible to describe genetic expression
data sets using a simple linear interaction model with only a small number of
interactions.  One important implication is that it is impossible to determine
the exact interactions among individual genes in these data sets.
The problem is underdetermined because the number of genes
is much larger than the number of time points in the experiments.
Nonetheless, we have shown that it is possible to accurately describe
the interactions among the characteristic modes.
Moreover, an interaction model with only two connections
reconstructs the key features of
the gene expression in the simplest cases with good fidelity.
Our results imply that because there are only a few essential 
connections among modes and therefore among genes, 
additional links provide redundancy in the
network. 

\vspace{1cm}

\noindent This work was supported by 
an Integrative Graduate Education and Research Training grant
from the National Science Foundation,
Istituto Nazionale di Fisica Nucleare (Italy), 
Komitet Badan Naukowych Grant 2P03B--146--18, 
Ministero dell' Universita e della Ricerca Scientifica,
National Aeronautics and Space Administration, 
and National Science Foundation Plant Genome Research
Program grant DBI-9872629.

\newpage

\begin{center}
TABLE 1\\
\end{center}

\noindent a) (cdc15)
$$ M =
\left [
\begin{array}{rrrrrrrrrrr}
 0.468 & -1.032 &   0.114 &  -0.199 &  -0.046 &   0.158 &   0.342 &  -0.360 &  -0.024 &   0.264 &  -0.519 \\ 
 0.695 &  0.517 &   0.007 &  -0.551 &  -0.011 &  -0.330 &  -0.183 &  -0.078 &  -0.175 &   0.190 &  -0.459 \\ 
 0.125 &  0.065 &   0.482 &   0.811 &  -0.105 &   0.027 &   0.165 &   0.153 &   0.008 &  -0.543 &   0.212 \\ 
-0.015 & -0.030 &  -0.182 &   0.306 &   0.543 &  -0.087 &   0.360 &  -1.113 &  -0.680 &  -0.993 &  -0.073 \\ 
 0.045 & -0.004 &  -0.339 &   0.225 &   0.498 &   0.433 &  -0.304 &   0.276 &   0.237 &   0.155 &  -0.223 \\ 
 0.007 &  0.027 &  -0.252 &  -0.017 &  -0.120 &  -0.321 &   0.628 &  -0.159 &   0.420 &   0.195 &   0.336 \\ 
 0.002 & -0.034 &  -0.104 &   0.061 &   0.005 &  -0.366 &  -0.299 &   0.145 &  -0.839 &   0.317 &   0.482 \\ 
 0.010 &  0.041 &  -0.030 &   0.053 &  -0.370 &   0.394 &  -0.175 &  -0.558 &  -0.093 &   0.559 &   0.075 \\ 
 0.016 & -0.005 &  -0.112 &  -0.032 &  -0.214 &   0.355 &   0.254 &   0.291 &  -0.349 &  -0.499 &   0.299 \\ 
 0.011 & -0.022 &  -0.087 &  -0.009 &  -0.200 &  -0.139 &  -0.426 &  -0.111 &   0.310 &  -0.535 &   0.161 \\ 
-0.019 &  0.002 &  -0.075 &   0.057 &  -0.192 &  -0.105 &   0.030 &   0.069 &  -0.185 &  -0.071 &  -0.840
\end{array}
\right ]
$$

\noindent b) (sporulation)
$$M =
\left [
\begin{array}{rrrrrr}
0.975  & -0.366 & -0.431 & -0.140 & -0.076 & 0.143 \\
0.096  & 0.734  & -0.636 & -0.186 & -0.032 & -0.143 \\
-0.223 & -0.386 & -0.090 & -0.650 & -0.482 & -0.417 \\
-0.086 & -0.059 & -0.396 &  0.587 & -0.482 & -0.046 \\
0.098  & -0.009 & -0.165 &  0.640 & 1.223 & 0.336 \\
0.002  & 0.035  &  0.590 &  0.182 &-0.576 & -0.965
\end{array}
\right ]
$$

\noindent c) (fibroblast)
$$M =
\left [
\begin{array}{rrrrrrrrrrrr}
0.760 & 0.313 & 0.334 & -0.116 & -0.732 & -1.389 & -0.954 & -0.456 & 0.199 & 0.290 & -0.341 & -1.661 \\
0.427 & 0.508 & -0.525 & 0.884 & 0.783 & 0.142 & 1.880 & -0.517 & -0.155 & -0.678 & 2.303 & 1.914 \\
-0.091 & 0.483 & 0.884 & -0.199 & -0.207 & 1.332 & -1.023 & -0.359 & -1.834 & 0.653 & -1.529 & 1.008 \\
-0.113 & 0.251 & 0.014 & 0.055 & 0.253 & 0.840 & -1.024 & 0.779 & -0.263 & 0.221 & -1.481 & -0.758 \\
0.012 & -0.057 & 0.525 & -0.317 & 0.281 & 0.820 & -0.051 & 0.284 & -0.422 & 0.274 & -1.249 & 0.191 \\
0.042 & 0.157 & 0.303 & -0.317 & -0.415 & 0.509 & -0.219 & -0.722 & -0.067 & -0.002 & -0.396 & -0.412 \\
-0.019 & 0.074 & 0.092 & -0.724 & -0.665 & -0.192 & 0.478 & -0.076 & 0.542 & -0.333 & -0.079 & 1.485 \\
0.114 & 0.085 & -0.108 & 0.183 & -0.187 & 0.510 & -0.109 & 0.165 & -0.349 & 0.256 & -0.020 & 1.381 \\ 
0.074 & 0.081 & 0.300 & -0.435 & -0.122 & -0.048 & -0.187 & -0.789 & -0.054 & -0.280 & -0.478 & 1.061 \\
-0.132 & -0.154 & -0.101 & 0.119 & 0.163 & -0.859 & 0.044 & -0.289 & 1.998 & 0.004 & -0.476 & -0.060 \\
0.057 & 0.044 & 0.155 & -0.091 & 0.038 & 0.383 & -0.148 & -0.447 & -0.343 & 0.139 & 0.319 & 0.254 \\
-0.013 & -0.050 & 0.072 & 0.267 & -0.084 & 0.223 & -0.265 & 0.071 & -0.201 & 0.122 & 0.617 & -0.753
\end{array}
\right ]
$$

\newpage 

$\;\;\;$

\newpage
\vspace{0.5cm}
\begin{center}
TABLE 2\\
\end{center}
(sporulation groups)
$$N =
\left [
\begin{array}{rrrrrr}
2.233 & -3.570 & 0.182 & -1.722 & -0.440 & -0.655 \\
1.913 & -1.921 & 0.509 & 0.118 & -0.356 & -0.287 \\
-0.707 & 3.949 & 1.219 & 2.638 & 1.175 & 0.707 \\
-1.157 & 0.422 & -0.421 & -0.525 & -0.169 & -0.130 \\
-0.905 & 0.954 & -0.640 & -0.515 & 0.823 & -0.057 \\
-1.294 & 0.699 & 0.212 & -1.232 & 1.014 & 0.635
\end{array}
\right ]
$$

\vspace{0.5cm}
\begin{center}
TABLE 3
\end{center}

a) (cdc15)
$$ M = 
\left [ 
\begin{array}{rr}
0.469 & -1.283 \\
0.621 & 0.468
\end{array}
\right ]
$$

b) (sporulation)
$$ M = 
\left [ 
\begin{array}{rr}
1.078 & -0.342 \\
0.214 & 0.812
\end{array}
\right ]
$$

c) (fibroblast)
$$ M = 
\left [ 
\begin{array}{rr}
0.941 & -0.045 \\
0.110 & 1.033
\end{array}
\right ]
$$

\vspace{0.5cm}
\begin{center}
TABLE 4
\end{center}
$$M = \left [ \begin{array}{rr} a & b \\ c & d \end{array} \right ]$$
$$G = \sqrt{ad-bc}$$
$$A = \sqrt{-b/c}$$
$$\tau =  \frac{2 \pi \Delta t}{\cos^{-1}\left(\frac{a+d}{2G}\right)}$$
$$\phi = \cos ^{-1} \left( \frac{a-d}{2 \sqrt{-bc}} \right) $$

\vspace{0.5cm}
\begin{center}
TABLE 5
\end{center}

\begin{center}
\begin{tabular}{|l|r|r|r|r|}    \hline
   & \multicolumn{1}{c|}{$G$} 
   & \multicolumn{1}{c|}{$A$} 
   & \multicolumn{1}{c|}{$\tau$} 
   & \multicolumn{1}{c|}{$\phi$} \\ \hline
cdc15       & 1.008 & 1.437 & 115 minutes & $90.0^\circ$ \\ \hline
sporulation & 0.974 & 1.264 & 12.8 hours & $60.6^\circ$ \\ \hline
fibroblast  & 0.988 & 0.640 & 29.2 hours & $130.8^\circ$ \\ \hline
\end{tabular}
\end{center}

\newpage
\begin{center}
REFERENCES
\end{center}

\parindent 0.0cm

1. Pease, A.\ C., Solas, D., Sullivan, E.\ J., Cronin, M.\ T., Holmes, C.\ P. \& Fodor, S. P. (1994)
{\it Proc.\ Natl.\ Acad.\ Sci.\ USA} {\bf 91}, 5022--5026.

2. Schena, M., Shalon, D., Davis, R.\ W. \& Brown, P. O. (1995) {\it 
Science} {\bf
270}, 467--470.

3. Liang, S., Fuhrman, S., \& Somogyi, R. (1998) {\it Pac. \ Symp.
 \ Biocomput.} {\bf 3}, 18--29.

4. Akutsu, T., Miyano, S., \& Kuhara, S. (1999)  {\it Pac. \ Symp.
Biocomput.} {\bf 4}, 17--28.

5. Chen, T., He, H.\ L., \& Church, G. M. (1999) {\it Pac. \ Symp.
\ Biocomput.} {\bf 4}, 29--40.

6. Szallasi, Z. (1999)  {\it Pac. \ Symp. \ Biocomput.} {\bf 4}, 5--16.

7. Weaver, D.\ C., Workman, C.\ T., \& Stormo, G.\ D. (1999) 
{\it Pac. \ Symp. \ Biocomput.} {\bf 4}, 112--23.

8. D'haeseleer, P.\ D., Wen, X., Fuhrman, S., Somogyi, R. (1999)
{\it Pac. \ Symp.\ Biocomput.\ } {\bf 4}, 41--52.

9. van Someren, E.\ P., Wessels, L.\ F.\ A., \& Reinders, M.\ J.\ T. (2000)
in {\it Proceedings of the Eighth 
International Conference on Intelligent Systems
for Molecular Biology. }, AAAI Press, Menlo Park, California.

10. Hartwell, L. H., Hopfield, J.\ J., Leibler, S., \& Murray, A.W.
(1999) {\it Nature} {\bf 402}, C47--52.

11. Jeong, H., Tombor, B., Albert, R., Oltvai, Z.\ N., \& Barabasi,
A. L. (2000)  {\it Nature} {\bf 407}, 651--4.

{\it Nature} {\bf 406}, 378--82.
%
509--12.

12. McAdams, H.\ H. \& Arkin, A. (1998)  {\it Annu. \ Rev.\ Biophys. \
Biomol. \ Struct.} {\bf 27}, 199-224.

13. McAdams, H. \ H. \& Arkin, A. (1999)  {\it Trends Genet.} {\bf 15}, 
65--9.

14. Bhalla, U. \ S. \& Iyengar, R. (1999) {\it  Science} {\bf 283}, 
381--387.

15. Alon, U., Surette, M.\ G., Barkai, N., \& Leibler, S. (1999)
{\it Nature} {\bf 397}, 168--71.

16. Barkai, N. \& Leibler, S. (1997) {\it Nature} {\bf 387}, 913--917.

17. Becskei, A. \& Serrano, L. (2000) {\it Nature} {\bf 405}, 590-593.

18. Yi, T. M., Huang, Y., Simon, M. I., \& Doyle, J. (2000) 
{\it Proc. \ Natl. \ Acad. \ Sci. USA} {\bf 97}, 4649--53.

19. Eisen, M.\ B., Spellman, P. \ T., Brown, P.\ O., \& Botstein, D.
(1998){\it  Proc. \ Natl. \ Acad. \ Sci. USA} {\bf 95}, 14863--14868.

20. Spellman, P.\ T., Sherlock, G., Zhang, M.\ Q., Iyer, V. \ R.,
Anders, K., Eisen, M.\ B., Brown, P. \ O., Botstein, D., \& Futcher, B.
(1998)  {\it Mol. \ Biol.\ Cell} {\bf 9}, 3273--3297.

21. Chu, S., DeRisi, J., Eisen, M., Mulholland, J., Botstein, D.,
Brown, P. \ O., \& Herskowitz, I. (1998)  
{\it Science} {\bf 282}, 699--705.

22. Iyer, V. R., Eisen, M. B., Ross, D. T., Schuler, G., Moore, T.,
Lee, J. \ C.\ F., Trent, J. M., Staudt, L.\ M., Hudson, J., Jr., Boguski,
M. S., et al. (1999)  {\it Science} {\bf 283}, 83--87.

23. Getz G., Levine, E., \& Domany, E. (2000) {\it Proc. \ Natl. 
\ Acad.  \ Sci. USA}
{\bf 97}, 12079--12084.

24. Raychaudhuri, S.\ Stuart, J.\ M., Altman, R. (2000) {\it Pac.\
Symp. \ Biocomput.} {\bf 5}, 452--463.

25. Holter, N.\ S., Mitra, M., Maritan, A., Cieplak, M., Banavar, J.\ R. 
\& Fedoroff, N.\ V. (2000)
{\it Proc.\ Natl.\ Acad.\ Sci.\ USA} {\bf 97}, 8409--8414.

26. Alter, O., Brown, P.\ O. \& Botstein, D. (2000)
{\it Proc.\ Natl.\ Acad.\ Sci.\ USA} {\bf 97}, 10101--10106.

27. Press, W.\ H., Flannery, B.\ P., Teukolsky, S.\ A. \& Vetterling, 
W.\ T. (1992)
{\it Numerical Recipes in C; The art of Scientific Computing}
(Cambridge Univ. Press, Cambridge), pp. 444--455.



\newpage
\vspace{0.5 cm}
\begin{center}
FIGURE CAPTIONS
\end{center}

\vspace{0.25 cm}
Figure 1. A comparison of measured and calculated expression profiles. 
Average expression profiles for the six clusters of 
genes in the sporulation data set (14) are represented by
circles and the approximated values calculated using the
best-fit time translation matrix are shown as lines.  

\vspace{0.25 cm}
Figure 2.  The first two characteristic modes for the a) cdc15, 
b) sporulation and c) fibroblast data sets.
The circles correspond to the measured data and the lines
show the approximations based on the best-fit $2 \times 2$ time 
translation matrices. 

\vspace{0.25cm}
Figure 3.  A reconstruction of the expression profiles for the cdc15
(first three panels), sporulation (middle three panels),
and fibroblast (last three panels) data sets. For each set,
panel a shows the results obtained using 
the $2 \times 2$ time translation matrix to determine the temporal evolution
of the expression profiles from their initial values, panel b
shows expression levels expressed as linear combinations of just
the two top modes, whereas panel c shows the experimental data.

\newpage
\begin{figure}
\vspace*{0.5cm}
\epsfxsize=3.0in
\hspace*{-0.5cm}
\centerline{\epsffile{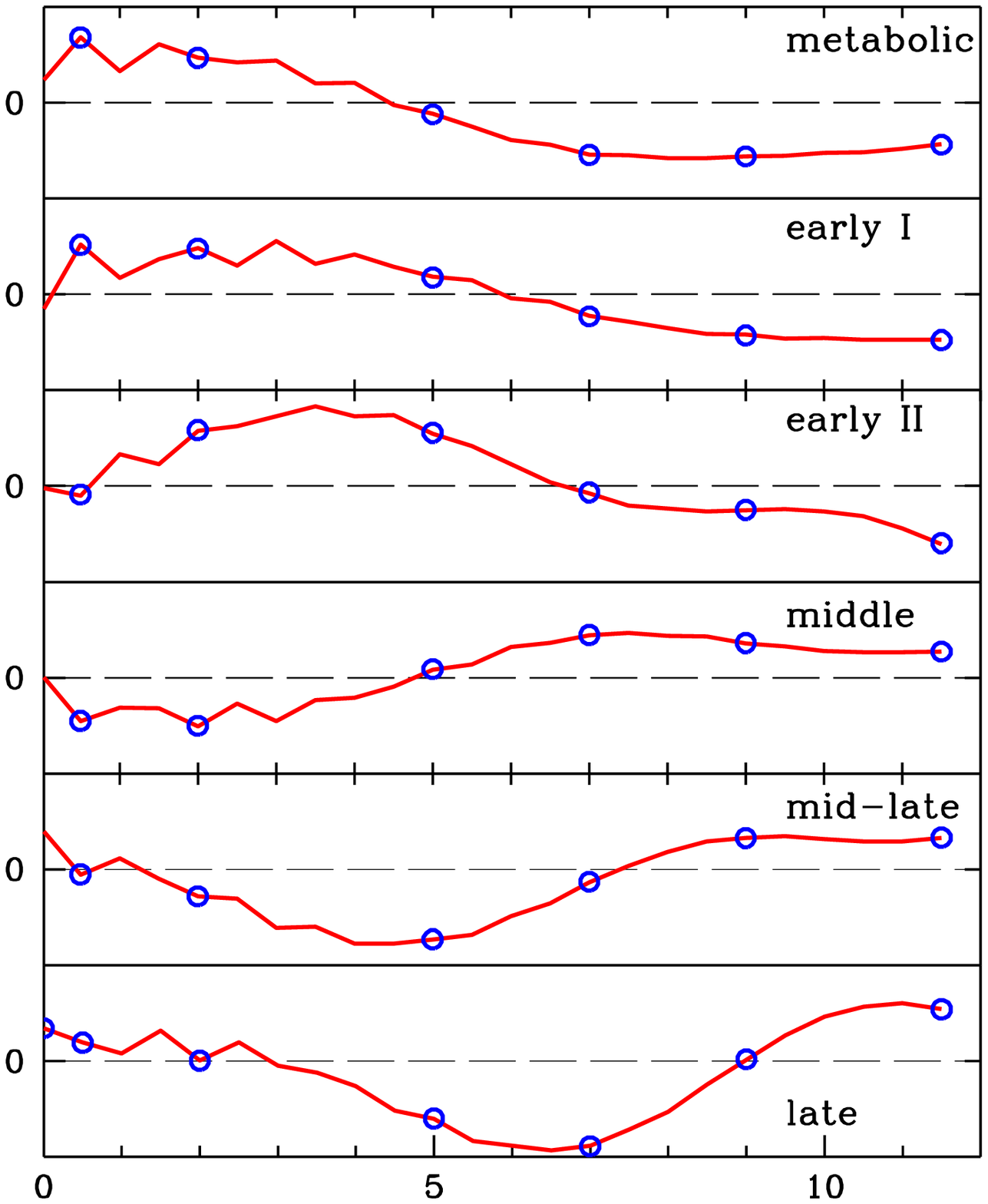}}
\vspace*{2.3cm}
\caption{ }
\end{figure}

\begin{figure}
\epsfxsize=3.8in
\centerline{\epsffile{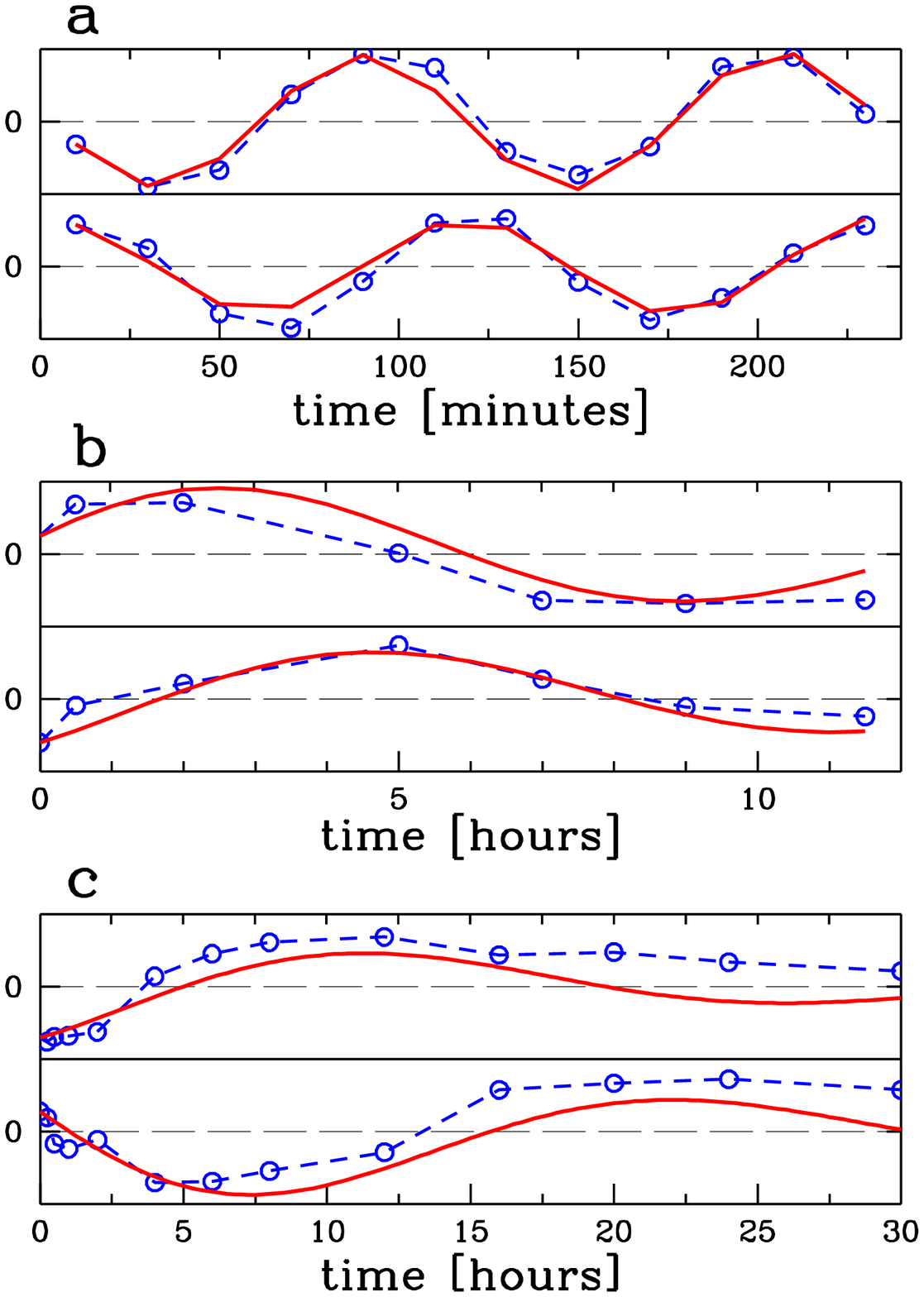}}
\vspace*{2.8cm}
\caption{ }
\end{figure}

\begin{figure}
\epsfxsize=3.8in
\centerline{\epsffile{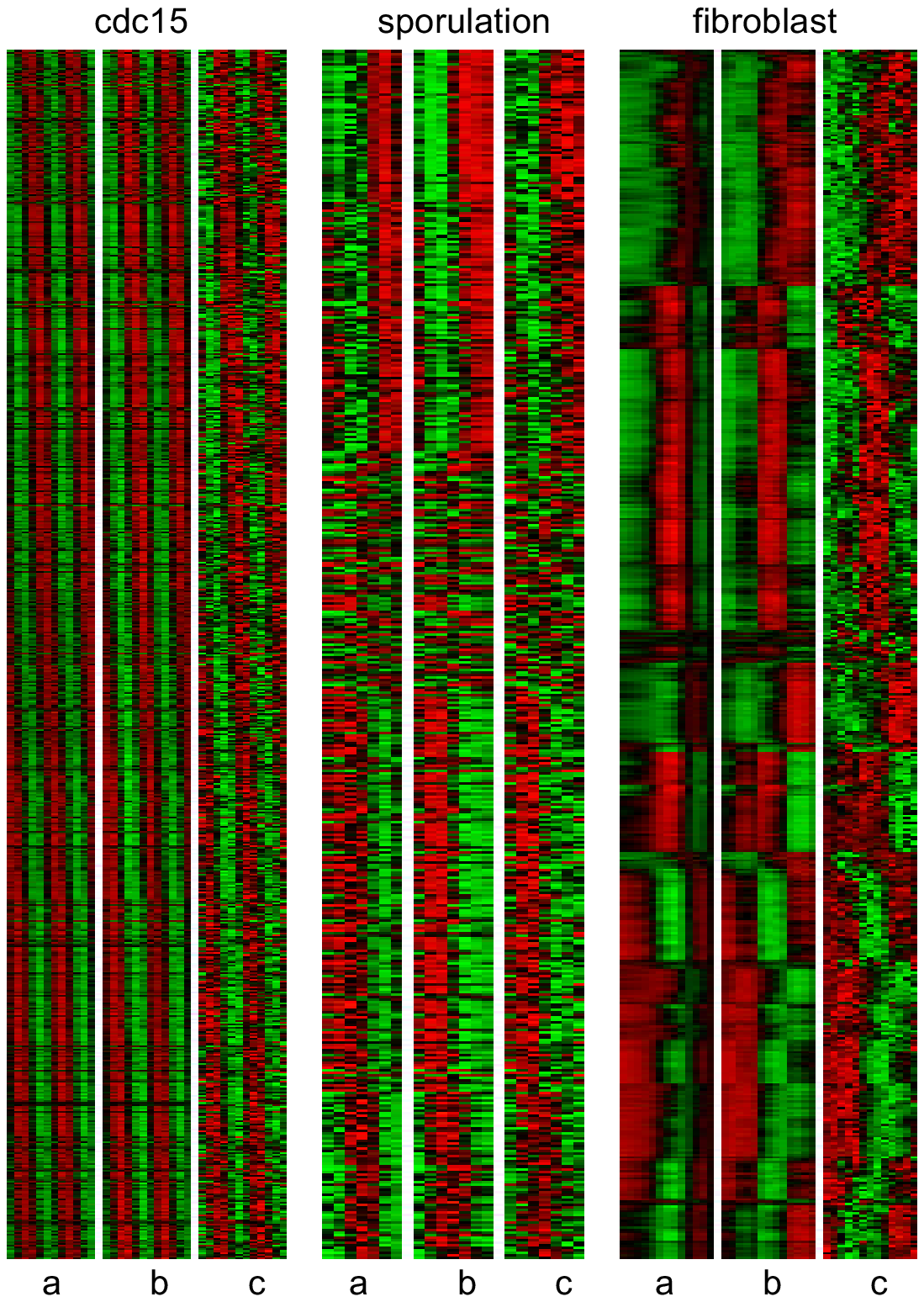}}
\vspace*{2.8cm}
\caption{ }
\end{figure}


\end{document}